\documentclass[prb,aps,twocolumn,showpacs]{revtex4}

\usepackage{epsfig}

\begin{document}

\title{Superconductivity and normal state properties of single-crystalline Tl$_{0.47}$Rb$_{0.34}$Fe$_{1.63}$Se$_2$ as seen via 
$^{77}$Se and $^{87}$Rb NMR}

\author{L. Ma ${^1}$}
\author{G. F. Ji ${^1}$}
\author{J. Zhang ${^2}$}
\author{J. B. He ${^1}$}
\author{D. M. Wang ${^1}$}
\author{G. F. Chen ${^1}$}
\author{Wei Bao ${^1}$}
\author{Weiqiang Yu ${^1}$}
\email{wqyu_phy@ruc.edu.cn}
\affiliation{
$^1$Department of Physics, Renmin University of China, Beijing 100872, China\\
$^2$School of Energy, Power and Mechanical Engineering, North China Electric Power University, Beijing 102206, China}
\date{\today}


\begin{abstract}

We report both $^{77}$Se and $^{87}$Rb NMR studies on Tl$_{0.47}$Rb$_{0.34}$Fe$_{1.63}$Se$_2$ single-crystalline superconductors  
($T_c\approx$ 32 K). Singlet superconductivity is decisively determined by a sharp drop of the Knight shift $K(T)$ below $T_c$, 
after subtracting the superconducting diamagnetic effect. However, the Hebel-Slichter coherence peak below $T_c$ is not observed 
in the spin-lattice relaxation rate $1/T_1$, even with a low in-plane NMR field of 2.6 Tesla. Just above $T_c$, no evidence of 
low-energy spin fluctuation is found in the spin-lattice relaxation rate on both the $^{77}$Se and the $^{87}$Rb sites. Upon 
warming, however, the Knight shifts and the spin-lattice relaxation rates of both nuclei increase substantially with temperature. 
In particular, the Knight shift is nearly isotropic and follow a function fit of $K=a+bT^2$ from $T_c$ up to 300 K. These normal 
state properties may be an indication of thermally activated spin fluctuations. Our observations should put a strong constraint to 
the theory of magnetism and superconductivity in the newly discovered iron-based superconductors.
   
\end{abstract}

\maketitle

\section{Introduction}
As the second type of high-temperature superconductors after cuprates, the iron-based superconductors \cite{Hosono_Jacs_130_3296, 
Chen_Nature_453_761, Ren_CPL_12_105, Chen_PRL_100_247002} have attracted a lot of research interests in recent years.
The iron pnictide superconductors are remarkable in that they originate from antiferromagnetic (AFM) semimetals 
\cite{Chen_PRL_100_247002}, and have multiple electron and hole bands on the Fermi surface which are gapped in the superconducting 
state \cite{Ding_EPL_83_47001, ZhouXJ_CPL}. Lately, a new family of iron selenide superconductors, with nominal chemical formulas 
$A$$_{y}$Fe$_{2-x}$Se$_2$ or (Tl, $A$)$_{y}$Fe$_{2-x}$Se$_2$ ($A$=K, Rb, Cs) \cite{Guo_PRB_82_180520, Mizuguchi_10124950, 
Chen_CM_10125525, FangMH_CM_10125236, Chen_CM_11010789}, have been discovered with a $T_c$ as high as 33 K. Comparing with iron 
pnictides, this new family show qualitatively different properties in the lattice structure, the band structure and the magnetic 
properties as described below, and thus enrich and challenge our understanding of the iron-based superconductivity. 

First, the composition of this type of superconductors is identified to be close to $A$$_{0.8}$Fe$_{1.6}$Se$_2$
by the x-ray and the neutron diffraction structure refinement studies \cite{BaoW_11014882, Bao_11020830}. This chemical 
stoichiometry leads to an almost perfect Fe-vacancy order in superconductors with a $\sqrt{5}$$\times$$\sqrt{5}$$\times$1 
supercell at low temperatures \cite{BaoW_11023674, LiJQ_CM_11012059, BaoW_11014882, BacsaJ_11020488,Bao_11020830}. Second, the 
Angle-resolved Photoemission Spectroscopy (ARPES) studies found that the hole bands centered at the $\Gamma$ point sink below the 
Fermi level \cite{Feng_CM_10125980,Ding_CM_10126017}, which is different from previous iron pnictide and chalcolgenide 
superconductors \cite{ZhouXJ_11014556, DingH_11014923, ZhouXJ_11021057}. Third, nodeless superconducting gaps are observed on the 
electron pockets on the Fermi surface \cite{Feng_CM_10125980, ZhouXJ_11014556, DingH_11014923, ZhouXJ_11021057}. NMR studies on 
nominal K$_{0.8}$Fe$_{2-x}$Se$_2$ show a singlet pairing symmetry from the Knight shift, whereas the Hebel-Slichter coherence peak 
is not observed in the spin-lattice relaxation rate \cite{YuW_11011017}. The $s^{\pm}$ gap symmetry, proposed in iron pnictides 
with interband transitions, may not be applicable here because of the absence of the hole band on the Fermi surface. Finally, an 
antiferromagnetic order with a large magnetic moments (2.3-3.3 $\mu _B$/Fe) has been determined by neutron diffraction 
\cite{Bao_11020830,Ye_11022882} for all of the new iron selenide superconductors, and possibly coexists with bulk 
superconductivity as indicated by the $\mu$SR and other bulk measurements \cite{Shermadini_11011873, Bao_11020830, 
ChenXH_11022783}. From NMR, however, the Curie-Weiss upturn in $1/T_1T$, as an indication of low-energy spin fluctuations observed 
in most iron pnictides \cite{Imai_prl_102_177005, Kita_JPSJ_77_114709,Baek_PRL, Yashima_JPSJ_NMR, Ning_PRL_104, Nakai_PRL_2010, 
Urbano_PRL, Ma_prb_2010}, is not seen in K$_{0.8}$Fe$_{2-x}$Se$_2$ \cite{YuW_11011017}. 

In the previous NMR studies, the superconducting diamagnetic effect is not counted, which may preclude a decisive evidence for a 
singlet pairing \cite{YuW_11011017, Kotegawa_11014572, ImaiT_11014967}. It should also be noted that the large NMR magnetic field 
used in those studies may suppress the Hebel-Slichter coherence peak. Furthermore, a large increase of the Knight shifts and the 
spin-lattice relaxation rates with temperature are also observed in the normal state with an unknown origin. So far NMR data are 
only available in K$_{0.8}$Fe$_{2-x}$Se$_2$. NMR studies in the same structure family are demanding to verify and understand these 
observations.

In this paper, we report both $^{77}$Se and $^{87}$Rb NMR studies on a single-crystalline Tl$_{0.47}$Rb$_{0.34}$Fe$_{1.63}$Se$_2$ 
superconductor. The paper is organized as following. First, we compare the Knight shifts of $^{77}$Se and $^{87}$Rb in the 
superconducting state. The superconducting diamagnetic effect are evaluated, thanks to the differences in the hyperfine coupling 
strength of the two nuclei. Second, we investigate the spin-lattice relaxation rate under different NMR fields below $T_c$. The 
coherence peak is not observed even with a 2.6 T in-plane field. Finally, we study the Knight shift and the spin-lattice 
relaxation rate in the normal state, and discuss the possible spin fluctuations. Detailed comparisons with 
K$_{0.8}$Fe$_{2-x}$Se$_2$ are also given in the paper.

\section{Experimental Techniques and Results}

\begin{figure}
\includegraphics[width=8cm, height=6cm]{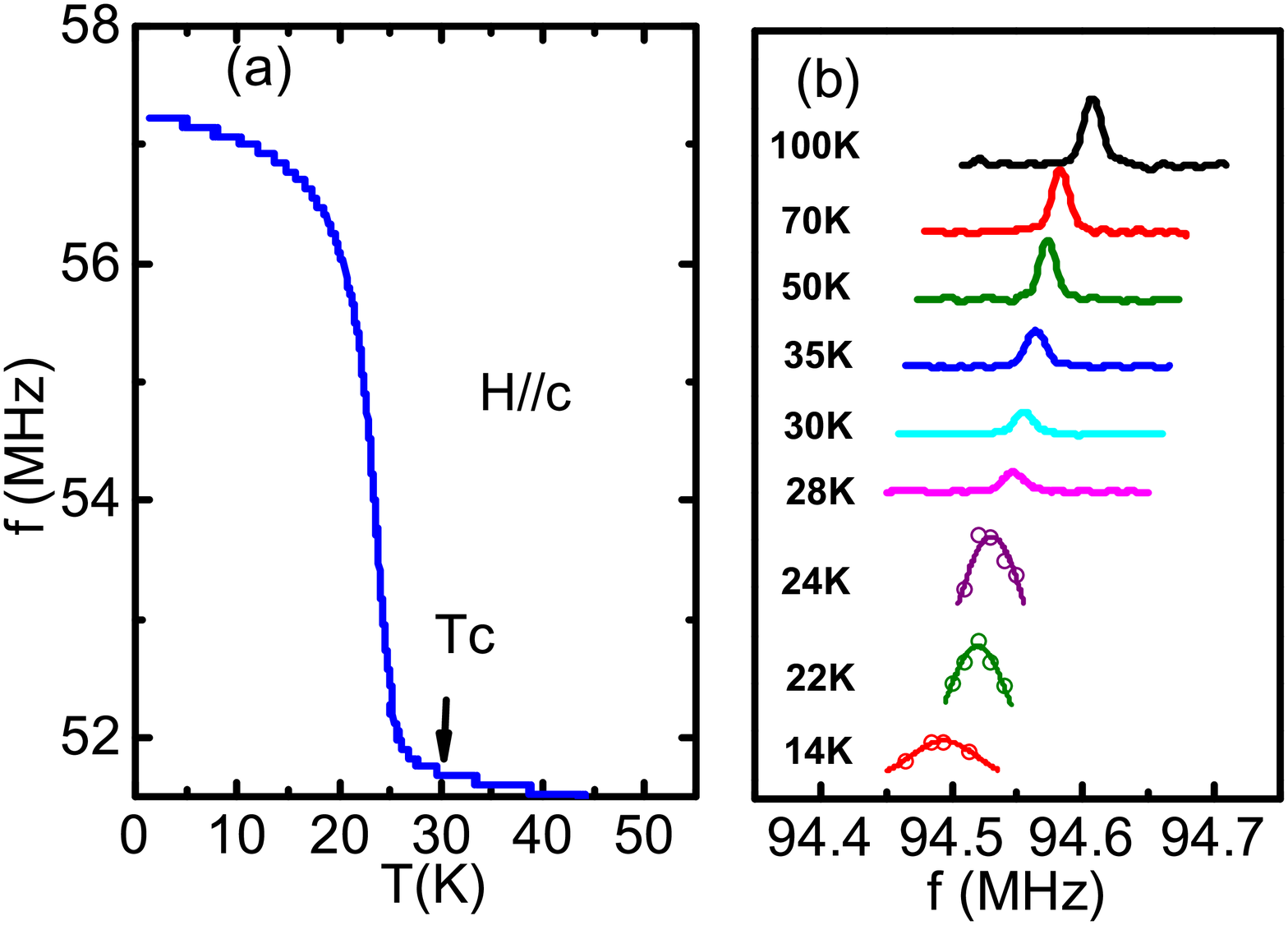}
\caption{\label{sus1}(color online) (a) The temperature shift of the tuning frequency $f$ of a fixed NMR circuit, with the field 
(11.62 Tesla) oriented along the crystalline $c$-axis. The large frequency shift below $T_c$ is related to the ac susceptibility 
$\chi_{ac}$ of the sample, $\Delta f \propto -\chi_{ac}$. $T_c$ of the sample is determined as $ 31\pm 1$ K with the applied 
field. (b) The $^{77}$Se NMR spectra at different temperatures. For data below 28 K, the circles represent the integrated spectral 
intensity, and the solid lines are Gaussian fit to the data. }
\end{figure}

The Tl$_{0.47}$Rb$_{0.34}$Fe$_{1.63}$Se$_2$ single crystals were synthesized by the Bridgeman method \cite{Chen_CM_11010789}. The  
chemical composition,  determined by the inductively coupled plasma (ICP) analysis, is consistent with the Fe-vacancy order in a 
general stoichiometric structure of $A_{0.8}$Fe$_{1.6}$Se$_2$ (or $A_{2}$Fe$_{4}$Se$_5$) \cite{BaoW_11023674}. In this paper, we 
primarily report our results on one single crystal with dimensions of $\sim$5$\times$3$\times$1mm$^3$ ($T_c$$\approx$$32 K$ at 
zero field). The superconducting transition was monitored {\it in situ} by the ac susceptibility measurements with the NMR coil. 
In Fig.~\ref{sus1}(a), $T_c$ is shown at about 31 K with an 11.62 T field oriented along the crystalline $c$ axis.

The $^{77}$Se ($S$=1/2) NMR is performed under 11.62 T NMR field, and $^{87}$Rb ($S$=3/2) NMR is performed under 11.62 T and 2.6 T 
field. We also studied the field anisotropy effect with field along both the crystalline $c$ axis and the $ab$ plane. Typical 
$^{77}$Se spectra with $H//c$ are shown in Fig.~\ref{sus1}(b). A linewidth about 20 kHz is observed at $T=$35 K, and narrows down 
as temperature increases. Below $T_c$, the spectra broaden and shift to lower frequencies. The $^{87}$Rb has one center transition  
and two satellites ($\nu _q\approx$1.4 MHz, data not shown).

The Knight shifts $^{77}K(T)$ and $^{87}K(T)$ are obtained from $K(T)=(f-\gamma _n B)/\gamma _n B$, where $f$ is the measured 
resonance frequency at magnetic field $B$, and $^{77}\gamma _n =8.131$ MHz/T  and $^{87}\gamma _n =13.931$ MHz/T are the 
gyromagnetic ratios of two respective nuclei. The Knight shift data of $^{77}$Se and $^{87}$Rb are shown in Fig.~\ref{knight2} and 
in Fig.~\ref{knightb}, respectively.  

The spin-lattice relaxation rate $1/T_1$ is measured by the inversion-recovery method, and is obtained from the fitting 
$I(t)/I(0)=1-ae^{-t/T1}$ for $^{77}$Se and $I(t)/I(0)=1-a(0.1e^{-t/T_1}-0.9e^{-6t/T1})$ for $^{87}$Rb ($a\ge 1$). The fitting 
works nicely at all measured temperatures, indicating a single phase. The $1/^{77}T_1$ and $1/^{87}T_1$ data are shown in 
Fig.~\ref{slrr3} and Fig.~\ref{slrrb}, respectively. 
   
\section{The NMR Knight shift and the singlet superconductivity} 
 
\begin{figure}
\includegraphics[width=9cm, height=6cm]{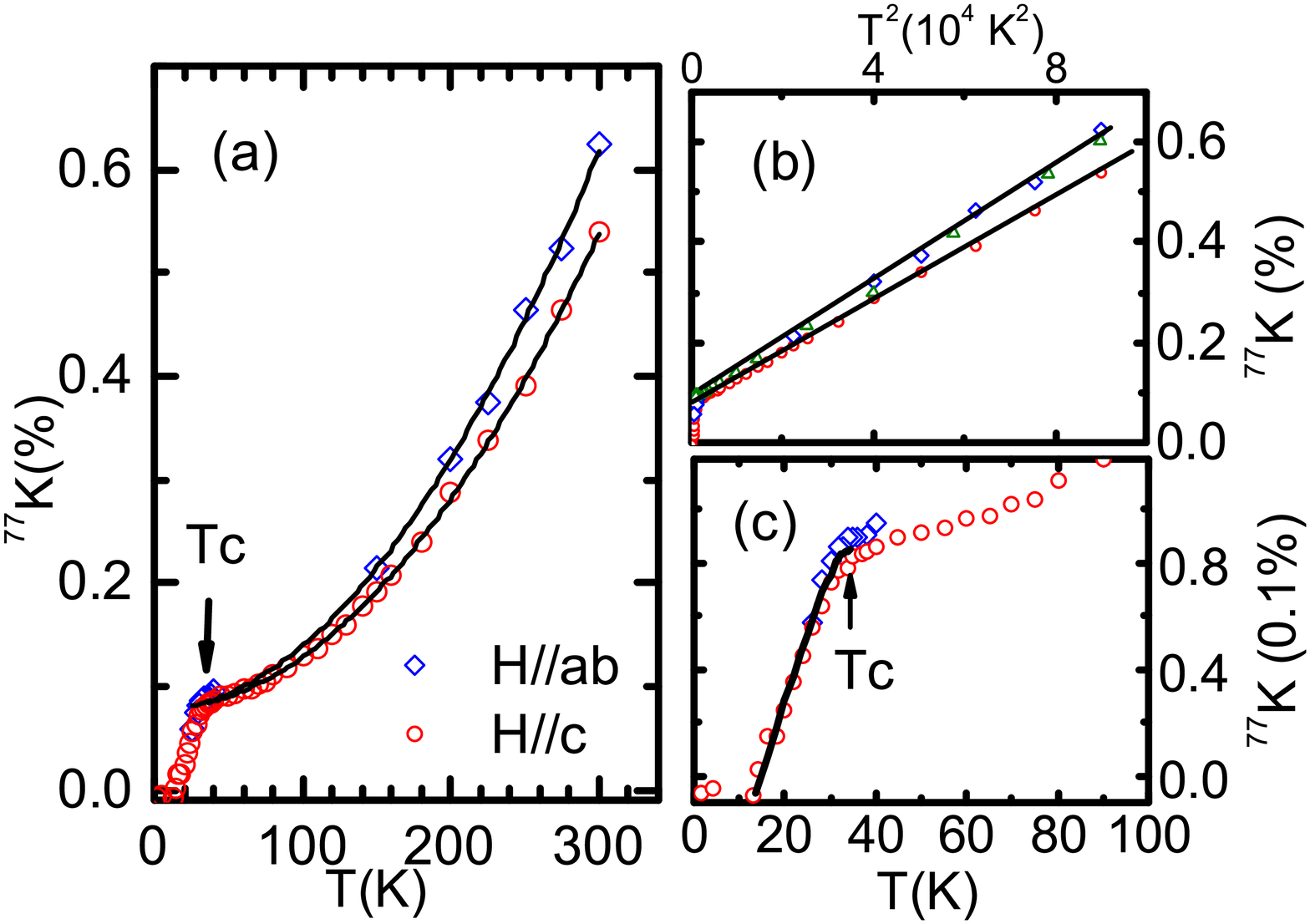}
\caption{\label{knight2}(color online) (a) The temperature dependence of the Knight shift $^{77}K$ under 11.62 T NMR field with 
$H//ab$ (hollow diamonds)  and $H//c$ (hollow circles). The solid lines are guides to the eye. (b) Replot of $^{77}K$ as a 
function of $T^2$ for both field orientations, and the solid lines are function fit to $K(T)=a+bT^2$. The up-triangles are data 
for K$_{0.8}$Fe$_{1-x}$Se$_2$ ($T_c\approx$32 K) with $H//ab$, adapted from Ref.~[\onlinecite{YuW_11011017}]. (c) The enlarged 
view of the low-temperature behavior of $^{77}K$ with two field orientations.} 
\end{figure}

The $^{77}$Se Knight shift data of Tl$_{0.47}$Rb$_{0.34}$Fe$_{1.63}$Se$_2$ are shown in Figure \ref{knight2}(a). The measurements 
were performed with field along either the $ab$-plane (hollow diamonds) or the $c$-axis (hollow circles). Above $T_c$, the 
$^{77}K$ for both field orientations increases monotonically with temperature. In fact, both sets of data can be nicely fit by a 
simple function $K_s(T)=a+bT^2$ with temperature from $T_c$ to 300 K, as shown in Fig.~\ref{knight2}(b). The fitting parameters 
are given as $a\approx 0.1\%$ and $b\approx 6\times 10^{-6}\%/$K$^2$ for $H//ab$, and $a\approx 0.1\%$ and $b\approx 5.5\times 
10^{-6}\%/$K$^2$ for $H//c$. The low-temperature part of $^{77}K(T)$ is highlighted in Figure \ref{knight2}(c) to show a fast drop 
of $^{77}K$ from $T_c$ to $T_c/2$ (about 15 K). Below 15 K, the spectral intensity is very small due to RF screening, and we 
believe our Knight shift is not well determined because that vortex core also contributes to the spectrum.

\begin{figure}
\includegraphics[width=9cm, height=6cm]{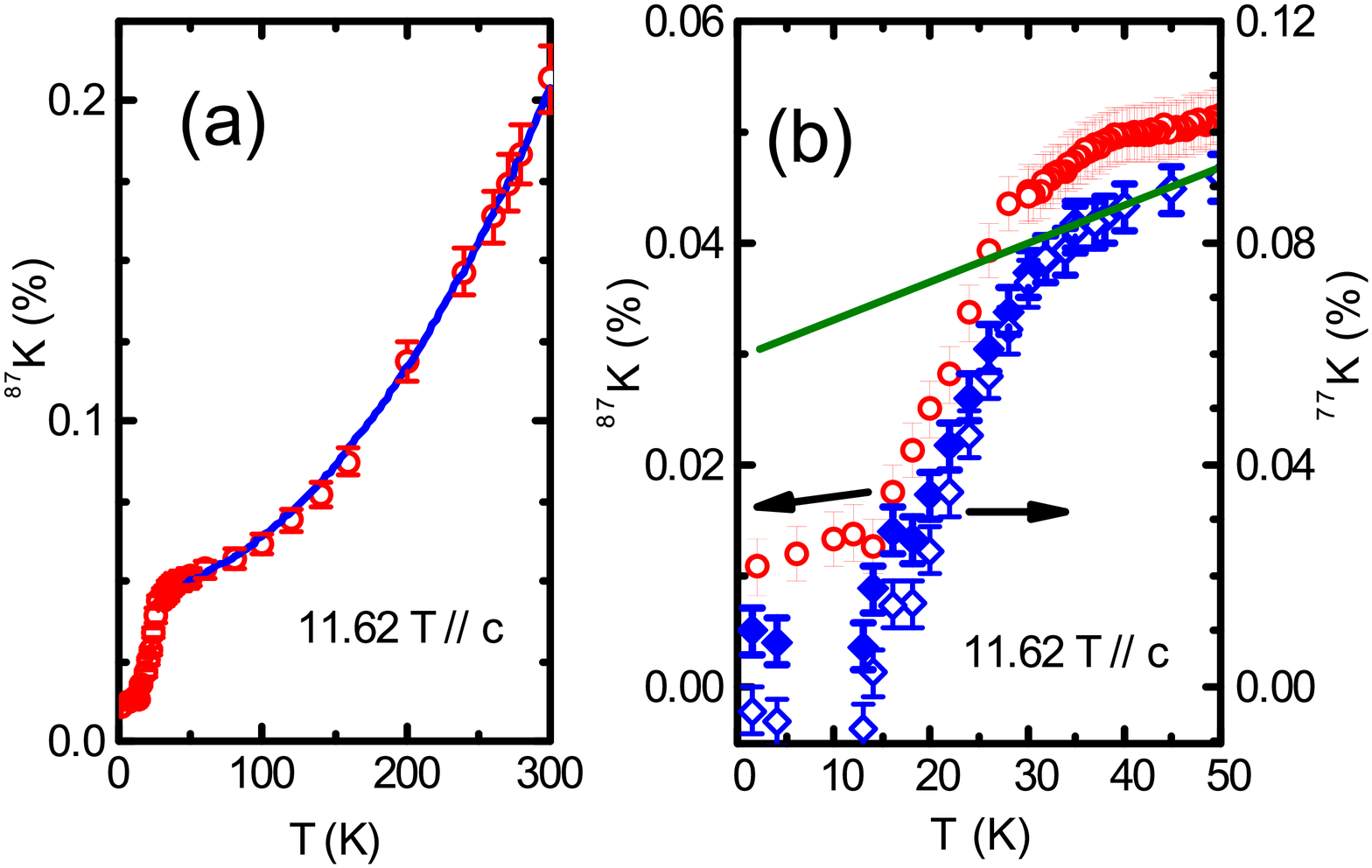}
\caption{\label{knightb}(color online) (a) The temperature dependence of the Knight shift $^{87}K$ under 11.62 T NMR field with 
$H//c$. The solid line is a function fit to $K(T)=a+bT^2$. (b) Replot of the low-temperature part of $^{87}K$ (hollow circles) and 
$^{77}K$ (hollow diamonds) under the same field condition. The solid line is a guide to the normal state $^{77}$K . The solid 
diamonds with error bar represent the values of $^{77}K$ after subtracting the superconducting diamagnetic effect (see text).}
\end{figure}

The $^{87}$Rb Knight shift data of Tl$_{0.47}$Rb$_{0.34}$Fe$_{1.63}$Se$_2$ are shown in Figure \ref{knightb}(a). The measurements 
were performed with field along the crystalline $c$-axis. Above $T_c$, the $^{87}K$ also increases with temperature, following the 
same $a+bT^2$ behavior (see the solid line fitting).

We first discuss the Knight shift data in the superconducting state and its implication to the pairing symmetry. Below $T_c$, the 
total shift $K$ includes three parts $K=$$K_d$$+$$K_s$$+$$K_c$, where $K_d$, $K_s$ and $K_c$ represent the superconducting 
diamagnetic shielding, the spin, and the chemical (or orbital) contribution to the frequency shift. In principle, $K_c$ does not 
change with temperature, while $K_d$ decrease from zero to negative below $T_c$, and $K_s$ measures local electron susceptibility. 
Therefore, below $T_c$, the change of the shift follows $\Delta$$K (T)$$=$$\Delta$$K_s(T)$$+$$\Delta K_d(T)$ with $\Delta K$ 
defined as $\Delta$$K$$ =$$K(T)$$-$$K(T_c)$. Usually, the $K_s$ and the $K_d$ are difficulty to separate. In the current compound, 
due to a weaker hyperfine coupling on the interlayer $^{87}$Rb site, as shown by the smaller Knight shift, $K_s$ and $K_d$ can be 
estimated by comparing the $^{77}K$ and the $^{87}K$ under the same field condition (see Fig.~\ref{knightb}(b)), as described 
below. 

Let's first look at the Knight shifts at a specific temperature $T$$=$16 K.  The $^{87}K$ decreases by about $0.028\pm 0.004\%$ 
from $T_c$ to 16 K, which gives the diamagnetic shift of $-0.028\%$$<$$\Delta$$K_d$$<$$0$. It is reasonable to assume that $K_d$ 
is the same for both nuclei from the same diamagnetic shielding, therefore the value of $^{77}K_s$$+$$^{77}K_c$ for $^{77}$Se 
should be moved up by the same amount $0.028\pm 0.004\%$ (or less) from the $^{77}K(T)$ ($0.015\pm 0.004\%$). This gives 
$^{77}K_s$$+$$^{77}K_c$$\le$$0.043\pm 0.008\%$ at 16 K, which is far below the guide line of the normal state $^{77}K$. 

The  $K_d$ and $K_s$ can also be calculated with a higher precision as following. We first estimate $^{87}K_s/^{77}K_s\approx 0.4$ 
from the change of the Knight shift in the temperature range between $T_c$ and 300 K, because the $^{77}K$ and the $^{87}K$ 
increases with the same $T^2$ behavior. Then the value of $K_d$ and $K_s$ are separated from the $^{77}K$ and the $^{87}K$ data at 
each temperature below $T_c$, using $^{87}K_s/^{77}K_s$$=$$0.4$. We plot the resulting value of $^{77}K_s$$+$$^{77}K_c$ below 
$T_c$ in Fig.~\ref{knightb}(b) (the solid diamonds). The data show that the diamagnetic shift is less than $0.02\%$ even at the 
lowest temperature.        

A sharp drop in $^{77}K$ just below $T_c$, after the subtraction of the diamagnetic effect, is clearly seen, indicating decisively 
a singlet superconductivity. This confirms the previous results in K$_{0.8}$Fe$_{2-x}$Se$_2$, although the diamagnetic effect was 
not considered there \cite{YuW_11011017}. The singlet pairing puts a strong constraint to the superconducting mechanism, and 
narrows down the gap to s-wave, d-wave, or other even orbital symmetry. 

\section{The spin-lattice relaxation rate and the coherence peak}

We further study the spin-lattice relaxation rate below $T_c$ to understand the pairing symmetry. Usually, the $1/T_1$ far below 
$T_c$ follows an activation behavior for a conventional $s$-wave superconductor. Unfortunately, the strong RF screening of our 
single crystal limits our studies of $T_1$ to a narrow temperature range below $T_c$. 

\begin{figure}
\includegraphics[width=8cm, height=7cm]{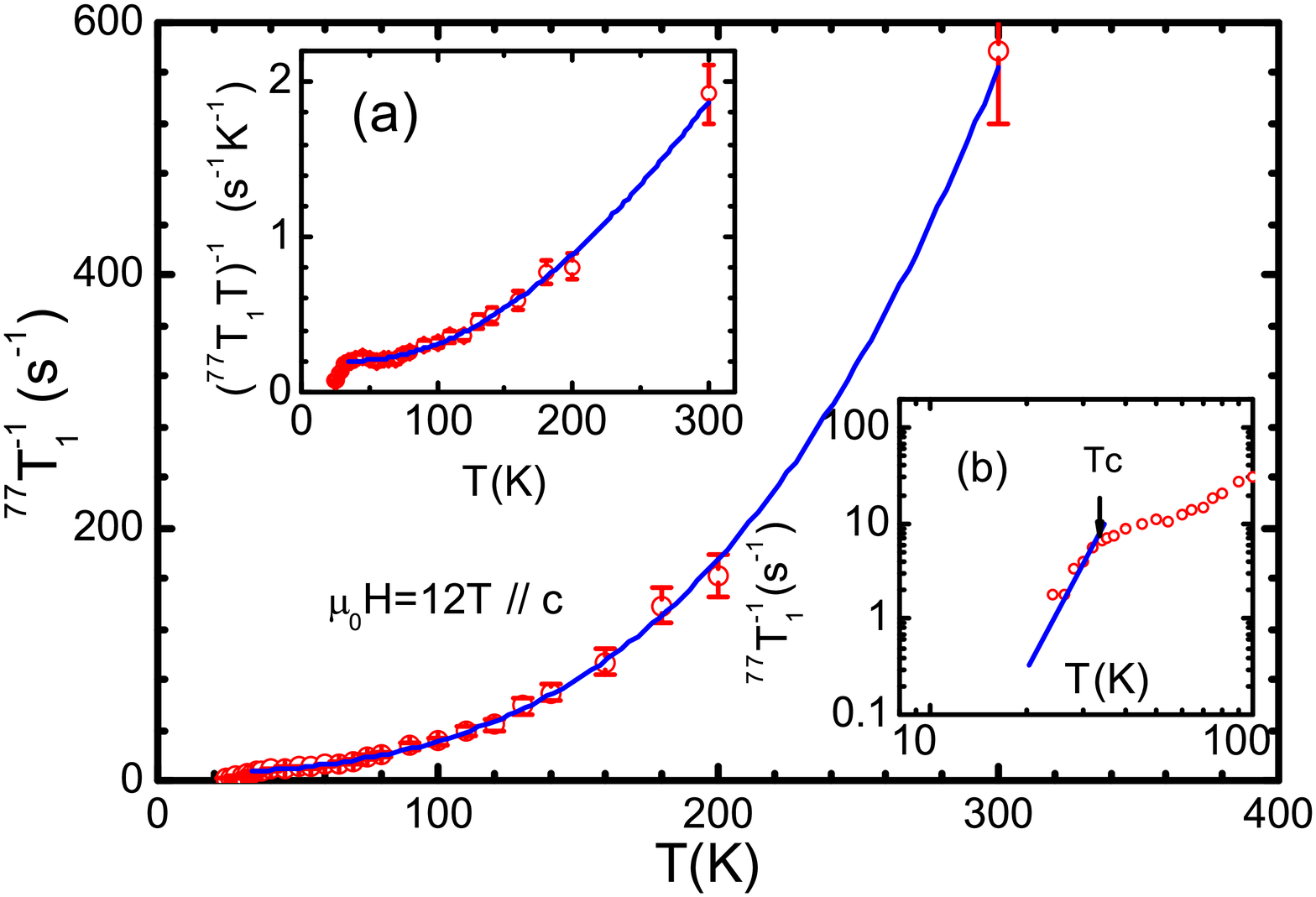}
\caption{\label{slrr3}(color online) The temperature dependence of the spin-lattice rate $1/^{77}T_1$ with 11.62 T field oriented 
along the $c$ axis. The solid line is a guide to the eye. Inset (a): the $1/^{77}T_1T$ vs. temperature; (b): The low-temperature 
$1/^{77}T_1$ data highlighted with a log-log scale. The solid line is a guide to the eye.}
\end{figure}

\begin{figure}
\includegraphics[width=9cm, height=7cm]{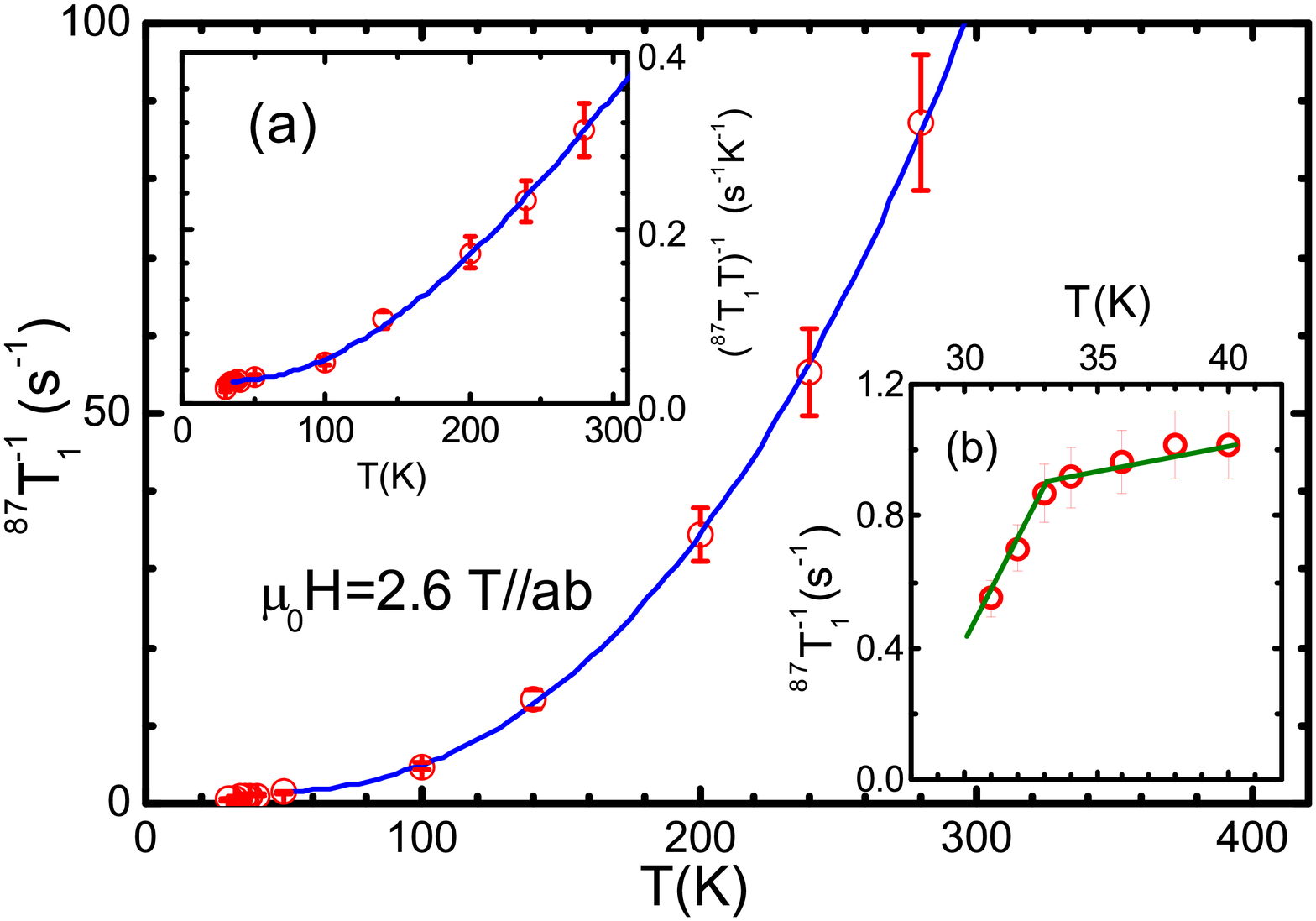}
\caption{\label{slrrb}(color online) The temperature dependence of the spin-lattice rate $1/^{87}T_1$ with 2.6 T field oriented 
along the $ab$ plane. The solid line is a guide to the eye. Inset (a): the $1/^{87}T_1T$ vs. temperature; (b): The highlighted 
low-temperature $1/^{87}T_1$ data. }
\end{figure}

The $1/^{77}T_1$ data with field oriented along the $c$ axis is shown in Fig.~\ref{slrr3}. Above $T_c$, the $1/^{77}T_1$ increases 
with  temperatures. Below $T_c$, a sharp drop of $1/^{77}T_1$ is clearly shown by the low-temperature highlight in 
Fig.~\ref{slrr3}(b). In Fig.~\ref{slrrb}, we show the $1/^{87}T_1$ data of $^{87}$Rb with a 2.6 T field oriented along the $ab$ 
plane. We are able to measure $1/^{87}T_1$ below $T_c$ due to the higher natural abundance of $^{87}$Rb. A sharp drop of 
$1/^{87}T_1$ is also prominent as shown in Fig.~\ref{slrrb}(b).

The above data show that the Hebel-Slichter coherence peak is absent in the $1/T_1$ just below $T_c$ for both nuclei. Since the 
$H_{c2}^{ab}(0)$ is estimated to be over 100 Tesla for the (Tl,Rb)$_{y}$Fe$_{2-x}$Se$_2$ superconductors \cite{FangMH_11010462}, 
the coherence peak is unlikely to be suppressed with the 2.6 T in-plane field. The absence of the coherence peak in the current 
compound is consistent with the result reported in K$_{0.8}$Fe$_{2-x}$Se$_2$ \cite{YuW_11011017, Kotegawa_11014572, 
ImaiT_11014967}.

Usually, the absence of coherence peak is interpreted as evidence for $d$-wave or other non $s$-wave pairing symmetry. In most 
iron pnictides, the coherence peak is not observed by far \cite{Grafe_PRL_101_047003, Matano_EPL_83_57001, Sato_JPSJ, 
Kobayashi_JPSJ_78_073704, Zhang_prb_81, Fukazawa_JPSJ_78_033704, Hammerath_CM_09123681}, which is proposed as a consequence of the 
$s^{\pm}$ pairing symmetry with interband impurity scattering \cite{Parker_PRB_78_134524, Chubukov_PRB_78_134512, 
Parish_PRB_78_144514, Bang_PRB_79_054529,Nagai_08091197}. In the new iron selenide superconductors, however, the $s^{\pm}$ 
scenario with interband transitions may not be applicable, because the hole band is absent on the Fermi surface close to the 
$\Gamma$ point. The absence of the coherence peak suggest that the system is different from a conventional $s$-wave superconductor 
and needs to be further addressed. It may be worthwhile to note that an $s^{\pm}$ symmetry is proposed in 
K$_{0.8}$Fe$_{2-x}$Se$_2$ considering the coupling between the bonding and the antibonding states on the Fermi surface 
\cite{Mazin_11023655}.

\section{The low-temperature spin dynamics}
  
Next we discuss the magnetic properties from the Knight shift and the spin-lattice relaxation rate data in the normal state. In 
Fig.~\ref{slrr3}(a), the $1/^{77}T_1T$ is plotted versus temperature to compare with the Fermi liquid behavior ($1/T_1T=const.$). 
Our data show that $1/^{77}T_1T$ reaches a constant value when the temperature is close to $T_c$. $1/^{87}T_1T$ also shows a 
similar temperature dependence as seen in \ref{slrrb}(a). The Knight shift data, $^{77}K$ and the $^{87}K$, also level off toward 
$T_c$ (see Fig.~\ref{knight2}(a) and Fig.~\ref{knightb}(a)). The level off of both the Knight shift and the $1/T_1T$ suggest that 
the system approaches a normal Fermi liquid behavior at low temperatures, with a constant density of states at the Fermi level.

In contrast, the binary iron selenide (FeSe) superconductors show strong antiferromagnetic spin fluctuations by a Curie-Weiss 
upturn in $1/T_1T\sim 1/(T+\Theta)$ \cite{Imai_prl_102_177005} above $T_c$. Similarly, the Curie-Weiss upturn in $1/T_1T$ has also 
been reported in many iron arsenide superconductors \cite{Kita_JPSJ_77_114709,Baek_PRL, Yashima_JPSJ_NMR, Ning_PRL_104, 
Nakai_PRL_2010, Urbano_PRL, Ma_prb_2010}. The existence of antiferromagnetic spin fluctuations just above $T_c$ in the previous 
iron-based superconductors draw a correlation between spin fluctuations and superconductivity \cite{Imai_prl_102_177005, 
Kita_JPSJ_77_114709,Baek_PRL, Yashima_JPSJ_NMR, Ning_PRL_104, Nakai_PRL_2010, Urbano_PRL, Ma_prb_2010}. 

The absence of the Curie-Weiss upturn in $1/^{77}T_1T$ in the current material and also in K$_{0.8}$Fe$_{2-x}$Se$_2$ 
\cite{YuW_11011017} suggests that the spin fluctuations are weak in both compounds, although their $T_c$ are also high. These 
distinctive behaviors of the spin-lattice relaxation rate in the new iron selenides does not seem to support that the 
superconductivity has a magnetic origin. 

However, we should be aware that spin fluctuations may exist in other forms. For example, some spin fluctuations at the wave 
vector $(\pm\pi, \pm\pi)$ cannot be detected in the current spin-lattice relaxation rate, due to the cancellation effect on the Se 
and the Rb sites which are located at the center of the Fe square. Furthermore, as we show below, the spin fluctuations may exist 
in the high-energy modes according to our high-temperature data. 
 
\section{The high-temperature spin dynamics} 

\begin{figure}
\includegraphics[width=9cm, height=7cm]{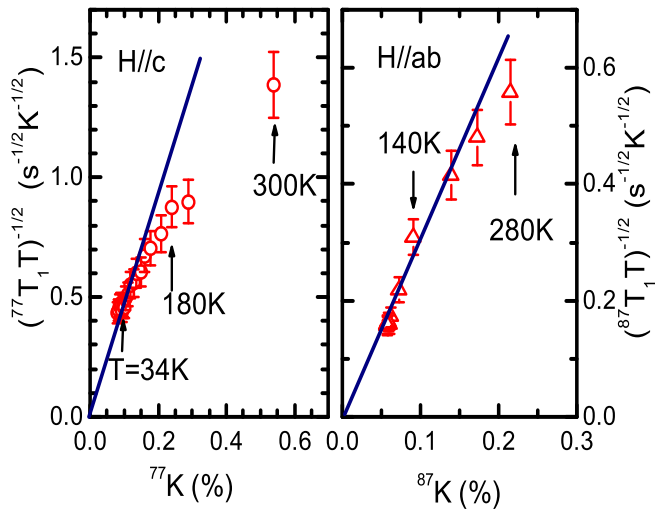}
\caption{\label{korringa5}(color online) The plot of $(^{77}T_1T)^{-0.5}$ vs. $^{77}K(T)$ with temperature as an implicit 
parameter. The solid line represents the Korringa relation. }
\end{figure}

To further compare with a Fermi liquid behavior, we plot $(T_1T)^{-0.5}$ against $K(T)$ in Fig.~\ref{korringa5}(a) and (b) for 
both nuclei. Usually, a Fermi liquid follows the Korringa relation $(T_1T)^{-0.5}\propto K_s$. As shown in 
Fig.~\ref{korringa5}(a), a linear relation between $(T_1T)^{-0.5}$ and $K(T)$, as demanded by the Korringa relation, is 
approximately satisfied for $^{77}$Se in the low-temperature limit. However, the Korringa relation is not obeyed in the high 
temperature range, indicating a deviation from the Fermi liquid behavior. Here we estimated $K_c\approx 0$ based on the value of 
$K$ at $T\ll T_c$, where $K_s(T)=0$ is expected (see Fig.~\ref{knightb}(b)). For $^{87}$Rb, the Korringa relation does not hold 
either at high temperatures.

Now we look closely on the high-temperature behavior of the Knight shift and the spin-lattice relaxation rate. The $1/T_1T$ of two 
nuclei increases substantially with temperature. The Knight shift data $^{77}K$ and the $^{87}K$ increase by over a factor of 
three, when the temperature increases from $T_c$ to the room temperature, as shown in Fig.~\ref{knight2} and Fig.~\ref{knightb}.  
The Knight shift $^{77}K$ of K$_{0.8}$Fe$_{2-x}$Se$_2$ with $H//ab$ is also plotted in Fig.~\ref{knight2} (b), and it falls on the 
same line of the current compound. This suggests a universal electronic structure in the new iron selenide superconductors. In 
contrast, in most iron pnictides, the Knight shift shows a much smaller increase with a linear-$T$ behavior \cite{Ning_PRL_104}.

The anisotropy of the Knight shift is small with  $K_s^{ab}/K_s^c\approx 1.1$ in the studied temperature range, as indicated by 
the slightly different slope in the $T^2$ plot of $^{77}K$ (see Fig.~\ref{knight2}). This anisotropy may come from an anisotropic 
susceptibility or an anisotropic Hyperfine coupling. As a local probe, the Knight shift data should indicate an intrinsic behavior 
of the magnetic susceptibility. In contrast, the magnetic susceptibility reported on Tl$_{0.58}$Rb$_{0.42}$Fe$_{1.72}$Se$_2$ shows 
a $T^2$ behavior with $H\parallel c$, but not with $H\parallel ab$ \cite{FangMH_11010462}, which may be affected by some unknown 
contributions. Furthermore, the ratio $^{87}K/^{77}K\approx 0.4$ indicates that the hyperfine field on the interlayer Rb site is 
comparable to that on the Se site, implying a three-dimensional like system. 

Such a large increase of the Knight shift is unlikely caused by charge excitations. In this compound, the charge excitation 
scenario will require an increase of the electron density over a factor of three up to the room temperature, which has not been 
indicated by any ARPES or transport measurements. Spin fluctuations is another possible scenario. The temperature behavior of the 
Knight shift suggest that a finite spin gap (or pseudogap) may exist, so that the spin fluctuations are gapped out at the low 
temperatures and thermally activated at high temperatures. However, the experimental evidence for the physical origin of such a 
spin gap is lacking. Our evidence of nearly isotropic, three-dimensional-like Knight shift should be important constraints to 
possible theories.   
 
We note that a magnetic transition at about 500 K has been reported by recent neutron scattering \cite{Ye_11022882} and 
susceptibility measurements \cite{ChenXH_11022783} on (Tl, $A$)$_y$Fe$_{2-x}$Se$_2$ superconductors. Unfortunately, the long-range 
antiferromagnetism has not been resolved by NMR so far. Further NMR measurements are demanding to address these issues.

\section{Summary and acknowledgment} 

To summarize, our $^{77}$Se and $^{87}$Rb Knight shifts of Tl$_{0.47}$Rb$_{0.34}$Fe$_{1.63}$Se$_2$ show a decisive bulk evidence 
for singlet superconductivity. However, the Hebel-Slichter coherence peak is not seen in the spin-lattice relaxation rate at very 
low in-plane field, which suggests that the system is probably not a conventional $s$-wave superconductor. In the normal state, 
evidence for low-energy spin fluctuations is not observed close to $T_c$, which questions the correlation between spin 
fluctuations and the superconductivity. With increasing temperature, however, the Knight shift increases substantially with a 
$T^2$ behavior up to room temperature, which is in contrast to most iron pnictides. The spin-lattice relaxation rate also shows a 
large increase with temperature. The Knight shift and the spin-lattice relaxation may suggest spin excitations with a finite gap 
(or pseudogap), which is thermally excited at high temperatures. 
 
Our resolved chemical composition of Tl$_{0.47}$Rb$_{0.34}$Fe$_{1.63}$Se$_2$ suggests that the $A_2$Fe$_4$Se$_5$ stoichiometric  
structure is formed in our superconducting compound. Furthermore, the similar NMR data of Tl$_{0.47}$Rb$_{0.34}$Fe$_{1.63}$Se$_2$ 
and K$_{0.8}$Fe$_{2-x}$Se$_2$ indicates the same electronic structure and pairing symmetry in both compounds, which points to a 
common origin of superconductivity in this new structure family of the iron-based superconductors.
 
The authors acknowledge Y. Su, Q. M. Zhang and X. J. Zhou for discussions on theoretical and experimental results, and Z. Y. Lu, 
T. Xiang and G. Zhang for pointing out the $T^2$ dependence of the Knight shift. Work at the RUC is supported by the NSFC (Grant 
No. 10974254, 11034012, and 11074304) and the National Basic Research Program of China (Grant No. 2010CB923000 and 2011CBA00100). 
JZ is supported by the Fundamental Research Funds for the Central Universities.    


\end{document}